\documentclass[twocolumn,aps,prb,superscriptaddress,10pt]{revtex4}%
\usepackage{graphicx}
\usepackage{amsmath}
\usepackage{txfonts}
\usepackage{amsfonts}
\usepackage{amssymb}%
\providecommand{\U}[1]{\protect\rule{.1in}{.1in}}

\begin{document}

\title{Electron tunneling through a single magnetic barrier in HgTe topological insulator}
\author{L. Z. Lin}
\affiliation{SKLSM, Institute of Semiconductors, Chinese Academy of
Sciences, Beijing, China}
\author{X. B. Wu}
\affiliation{Huaneng shandong shidao bay nuclear power co.,ltd, China}
\author{W. K. Lou}
\affiliation{SKLSM, Institute of Semiconductors, Chinese Academy of
Sciences, Beijing, China}
\author{D. Zhang}
\affiliation{SKLSM, Institute of Semiconductors, Chinese Academy of
Sciences, Beijing, China}
\author{Zhenhua Wu}
\affiliation{Institute of Microelectronics, Chinese Academy of Sciences, Bejing, China}
\email{wuzhenhua@ime.ac.cn}

\begin{abstract}
Electron tunneling through a single magnetic barrier in a HgTe topological insulator has been theoretically investigated. We find that the perpendicular magnetic field would not lead to spin-flip of the edge states due to the conservation of the angular moment. By tuning the magnetic field and Fermi energy, the edge channels can be transited from switch-on states to switch-off states and the current can be transmitted from unpolarized states to totally spin polarized states. These features offer us and efficient way to control the topological edge state transport, and pave a way to construct the nanoelectronic devices utilizing the topological edge states.

\end{abstract}

 \maketitle

In topological insulators (TIs) the energy states are fundamentally modified from ordinary insulators by
strong spin-orbit interactions, giving rise to a topologically distinct state of matter with a gapped insulating bulk
and a gapless metallic edge or surface.\cite{Hasan2010} Topological states of matter are characterized by bulk invariant:
Chern numbers\cite{Thouless,Kohmoto} or by a $\mathbb{Z}_{2}$ invariant,\cite{Mele} depending on if the time reversal symmetry is broken or conserved. In the case of the $\mathbb{Z}_{2}$ insulator the edge states are formed by time-reversed modes, so-called Kramers' partners, and they are helical, \cite{Mele2005} i.e., Kramers' partners counterpropagate along a given edge of the sample.
Owing to the linear dispersion and topological invariant, various interesting phenomena,
including edge or surface transport of spin-filtered Dirac fermions that are immune to localization,
have been predicted and raised expectations for novel applications.\cite{MooreN,Qixl,Mele,Mele2005,Bernevig,Sheng,Ful,Moore,Zhu,Chang}
For example, in a two-dimensional HgTe/CdTe quantum well with an inverted band structure, the helical edge states have been demonstrated experimentally in the Hall bar geometry.\cite{Konig,Roth,Knez} The edge states of the quantum spin Hall (QSH) system show spin-momentum locking in the sense that right-moving (left-moving) electrons are strictly spin up (spin down).\cite{Konig2008} So density-density interactions cannot lead to backscattering process since they cannot flip the spin, but magnetic impurities or magnetic field can lead to backscattering. This offers great potential application for a new generation of spintronic devices for low-power information processing.\cite{Chang,Kchang} It is therefore of interest to study electronic transport in prototypical device geometries involving topological edge states (TESs).\cite{Zhang2010,Zhang2011,Chang}

\begin{figure}[tbp]
\centering
\includegraphics[width=0.9\columnwidth]{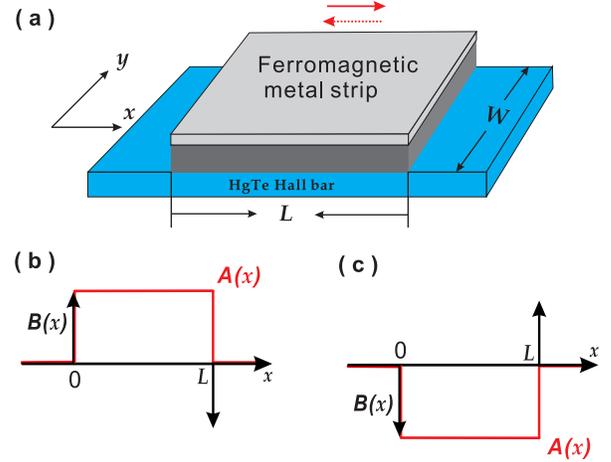}
\caption{(color online) (a) Schematic illustration of the considered structure
created by a ferromagnetic metallic strip on the top of the HgTe Hall bar.
The red horizontal arrows denote the magnetization directions of the strips that generate
a single delta-function-shaped magnetic barrier.
(b)/(c) Configuration for the delta-function-shaped barrier denoted by the solid/dashed red horizontal arrows as shown in Fig. \ref{fig:model}(a).}
\label{fig:model}
\end{figure}
In this work, we investigate theoretically the electron tunneling through a HgTe Hall bar
with an external perpendicular magnetic field, as shown in Fig. \ref{fig:model}(a).
We consider a delta-function-shaped magnetic barrier generated by the ferromagnetic metal strip on the top the HgTe Hall bar.
Such magnetic field profiles approximate those realized by depositing a patterned
superconducting plate above the HgTe Hall bar in the presence of a perpendicular magnetic field or by depositing one ferromagnetic
metal strip with the magnetization parallel to the surface on top of the HgTe Hall bar. Thus, we can create magnetic potentials
underneath the superconducting plate or the ferromagnetic strip.\cite{Ye_Nogaret,Matulis}
In the presence of an external magnetic field, the momentum is $\mathbf{P}$ = $\mathbf{p}$ + e$\mathbf{A}$,
where the vector potential $\mathbf{A}$ is generated by the magnetic metal strips.
We choose the Landau gauge in this work, i.e., $\mathbf{A}$ = $(0,A_y,0)$.
Electron states in the low-energy regime in the HgTe quantum well with inverted band structures can
be well described by a four-band effective Hamiltonian, including two $\Gamma_{6}$ electron bands $\left\vert {e\uparrow
}\right\rangle ,\left\vert {e\downarrow}\right\rangle $ and two $\Gamma_{8}$
heavy hole bands $\left\vert {hh\uparrow}\right\rangle ,\left\vert
{hh\downarrow}\right\rangle $. The total $4\times4$ Hamiltonian in the basis
$\left\vert {e\uparrow}\right\rangle ,\left\vert {hh\uparrow}\right\rangle
,\left\vert {e\downarrow}\right\rangle $, and $\left\vert {hh\downarrow
}\right\rangle $ is $\mathbf{H} = \mathbf{H_0} + \mathbf{H_Z}$,
with $\mathbf{H_0}$ given by\cite{Bernevig}
\begin{equation}
\mathbf{H_0}=\left(  {%
\begin{array}
[c]{cccc}%
{{\varepsilon_{\hat{\mathbf{k}}}}+M({\hat{\mathbf{k}}})} & {A{\hat{k}_{-}}} &
{0} & 0\\
{A{\hat{k}_{+}}} & {{\varepsilon_{\hat{\mathbf{k}}}}-M({\hat{\mathbf{k}}})} &
0 & 0\\
{0} & 0 & {{\varepsilon_{\hat{\mathbf{k}}}}+M({\hat{\mathbf{k}}})} &
{\ -A{\hat{k}_{+}}}\\
0 & 0 & {\ -A{\hat{k}_{-}}} & {{\varepsilon_{\hat{\mathbf{k}}}}-M({\hat
{\mathbf{k}}})}%
\end{array}
}\right),
\end{equation}
and $\mathbf{H_Z}$ is induced by Zeeman spin-spitting,
\begin{equation}
\mathbf{H_Z} = \left( {\begin{array}{*{20}{c}}
   {{g_E}} & 0 & 0 & 0  \\
   0 & {{g_{HH}}} & 0 & 0  \\
   0 & 0 & { - {g_E}} & 0  \\
   0 & 0 & 0 & { - {g_{HH}}}  \\
\end{array}} \right){\mu _B}B,
\end{equation}
where $\hat{\mathbf{k}}=({\hat{k}_{x}},{\hat{\Pi}_{y}})$ is the in-plane
momentum operator, ${\varepsilon_{\hat{\mathbf{k}}}}=C-D(\hat{k}_{x}^{2}%
+\hat{\Pi}_{y}^{2})$, $M(\hat{\mathbf{k}})=M-B(\hat{k}_{x}^{2}+\hat{\Pi}_{y}%
^{2}), {\hat{k}_{\pm}}={\hat{k}_{x}}\pm i{\hat{\Pi}_{y}}$, ${\hat \Pi _y} = {\hat k_y} + \frac{e}{\hbar }{A_y}$.
For a delta-function-shaped magnetic barrier case, ${A_y} = B \cdot L$ for $x \in [0,L]$ and $0$ otherwise.
$A,B,C,D$, and $M$ are the parameters describing the band
structure of the HgTe/CdTe quantum well. Note that the QSH and band insulator
states are distinguished by the different signs of the parameter $M$, which in
turn is determined by the thickness of the HgTe/CdTe quantum
well.\cite{Bernevig} ${g_E}$ and ${g_{HH}}$
denote the $g$-factor of the electron and heavy hole, respectively.
The velocity operator along the $x$ axis is given by
$\hat{v}_{x}=\partial\mathbf{H}(\hat{{k}_{x}},\hat{\Pi _y})/\partial\hat{k}_{x}$.

For a quasi-one-dimensional QSH bar system shown in Fig. 1(a), the transmission
and conductance is obtained by discretizing the quasi-one-dimensional system
into a series of transverse strips along the $x$ axis. In a given strip, the eigenstate can
be written as $\boldsymbol{\psi}(x,y)=e^{ik_{x}x}\boldsymbol{\phi}(y)$. The
Schr\"{o}dinger equation $E\boldsymbol{\psi}=\mathbf{H}(\hat{k}_{x}%
,\hat{\Pi _y})\boldsymbol{\psi}$ reduces to $E\boldsymbol{\phi
}(y)=\mathbf{H}(k_{x},\hat{\Pi _y})\boldsymbol{\phi}(y)$. The hard-wall boundary
conditions along the upper edge $y=W$ and lower edge $y=0$ enables the
expansion of $\boldsymbol{\phi}(y)$ using the complete basis $\varphi
_{n}(y)=\sqrt{2/W}\sin(n\pi y/W)$ $(n=1,2,\cdots,N_{\mathrm{cut}}$, where
$N_{\mathrm{cut}}$ is sufficiently large to ensure convergence$)$ as
$\boldsymbol{\phi}(y)=\sum_{n}\boldsymbol{\chi}_{n}\varphi_{n}(y)$. This gives%
\begin{equation}
E\boldsymbol{\chi}=\mathbf{H}(k_{x})\boldsymbol{\chi},\label{EIGEN_EQ}%
\end{equation}
where $\boldsymbol{\chi}\equiv\lbrack\boldsymbol{\chi}_{1},\cdots
,\boldsymbol{\chi}_{N_{\mathrm{cut}}}]^{T}$ is a $4N_{\mathrm{cut}}\times1$
vector and $\mathbf{H}(k_{x})$ is a $4N_{\mathrm{cut}}\times4N_{\mathrm{cut}}$
matrix with $\mathbf{H}_{m,n}(k_{x})\equiv\left\langle \varphi_{m}\left\vert
\mathbf{H}(k_{x},\hat{\Pi _y})\right\vert \varphi_{n}\right\rangle $. For a
given real $k_{x}$, there are $4N_{\mathrm{cut}}$ eigenenergies $E(k_{x})$ and
eigenvectors $\boldsymbol{\chi}(k_{x})$. Conversely, for a given real energy
$E$, there are $8N_{\mathrm{cut}}$ solutions $k_{x}(E)$ (which are in general
complex) and $\boldsymbol{\chi}(E)$. To find these solutions, we expand
$\mathbf{H}(k_{x})=\mathbf{H}^{(0)}+\mathbf{H}^{(1)}k_{x}+\mathbf{H}%
^{(2)}k_{x}^{2}$ into power of $k_{x}$ and define $\mathbf{F}\equiv
k_{x}\boldsymbol{\chi}$, then Eq. (\ref{EIGEN_EQ}) can be written as:%
\begin{equation}%
\begin{bmatrix}
0 & 1\\
E-\mathbf{H}^{(0)} & -\mathbf{H}^{(1)}%
\end{bmatrix}%
\begin{bmatrix}
\boldsymbol{\chi}\\
\mathbf{F}%
\end{bmatrix}
=k_{x}%
\begin{bmatrix}
1 & 0\\
0 & \mathbf{H}^{(2)}%
\end{bmatrix}%
\begin{bmatrix}
\boldsymbol{\chi}\\
\mathbf{F}%
\end{bmatrix}
.
\end{equation}
Solving the above $8N_{\mathrm{cut}}\times8N_{\mathrm{cut}}$ matrix equation
as a generalized eigen-problem gives $8N_{\mathrm{cut}}$ solutions for
$k_{x}(E)$ and $\boldsymbol{\chi}(E)$, corresponding to $8N_{\mathrm{cut}}$
eigenstates $\boldsymbol{\psi}(x,y)$ with energy $E$. These eigenstates are
classifed into the right-moving and left-moving states. The
right-moving states have a real $k_{x}$ and positive velocity $\left\langle
\boldsymbol{\psi}|\hat{v}_{x}|\boldsymbol{\psi}\right\rangle $ or have a
complex $k_{x}$ with $\operatorname{Im}k_{x}>0$. The left-moving states have a
real $k_{x}$ and negative velocity $\left\langle \boldsymbol{\psi}|\hat{v}%
_{x}|\boldsymbol{\psi}\right\rangle $ or have a complex $k_{x}$ with
$\operatorname{Im}k_{x}<0$. For a given energy $E$, the $m$th right-moving (left-moving)
states in the $j$th strip is denoted by $\boldsymbol{\psi}_{+,m}^{(j)}(E)$ ($\boldsymbol{\psi}_{-,m}^{(j)}(E)$).

Assuming that an electron with Fermi energy $E_{F}$ is injected into the $m$th
right-moving (with a real $k_{x}$) eigenstate $\boldsymbol{\psi}_{m,+}^{(L)}$
of the left lead, the scattering wave function in different strips of the Hall
bar structure is
\begin{align*}
\boldsymbol{\psi}^{(L)} &  =\boldsymbol{\psi}_{m,+}^{(L)}(E_{F})+\sum_{n}%
b_{n}^{(L)}(E_{F})\boldsymbol{\psi}_{n,-}^{(L)}(E_{F}),\\
\boldsymbol{\psi}^{(j)} &  =\sum_{n}a_{n}^{(j)}(E_{F})\boldsymbol{\psi}%
_{n,+}^{(j)}(E_{F})+\sum_{n}b_{n}^{(j)}(E_{F})\boldsymbol{\psi}_{n,-}%
^{(j)}(E_{F}),\\
\boldsymbol{\psi}^{(R)} &  =\sum_{n}a_{n}^{(R)}(E_{F})\boldsymbol{\psi}%
_{n,+}^{(R)}(E_{F}).
\end{align*}
By scattering matrix theory,\cite{Xu93,Lin2013} we obtain the total
conductance from the Landauer-B\"{u}ttiker formula,
\begin{equation}
G=G_{0}\sum\limits_{m\in L,n\in R}T_{n\leftarrow m}(E_{F})={G_{0}%
\sum\limits_{m\in L,n\in R}\frac{{v_{n}^{R}}}{{v_{m}^{L}}}}{\left\vert
{a_{n}^{R}}(E_{F})\right\vert ^{2}},
\end{equation}
where the sum runs over all right-moving modes in the left and right leads and
${G_{0}}={e^{2}}/h$ is the conductance unit.

\begin{figure}[tbp]
\centering
\includegraphics[width=0.95\columnwidth]{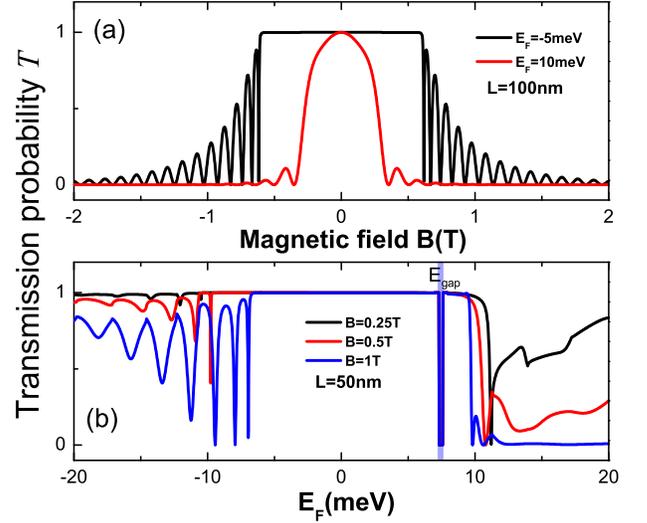}
\caption{(color online) The edge state transmission probability $T={T_{up}}={T_{down}}$
as a function of (a) the magnetic flux or (b) the Fermi energy $E_F$ without consideration
of the Zeeman term $\mathbf{H_Z}$.}
\label{fig:2}
\end{figure}
In this paper we focus on the control of the edge-state transport in the presence of the magnetic barrier. When $E_F$ lies in the bulk gap of the left and right leads, it only intersects the four edge channels in the leads. Therefore, we define ${T_{ \uparrow  \leftarrow  \uparrow }} = {T_{up}}$, ${T_{ \downarrow  \leftarrow  \downarrow }} = {T_{down}}$ and ${T_{ \downarrow  \leftarrow  \uparrow }} = {T_{ \uparrow  \leftarrow  \downarrow }} = {T_{\mathrm{sf}}}$. We find that the perpendicular magnetic field can't lead to spin-flip for edge states, i.e., $T_{\mathrm{sf}}=0$. That is because the momentum matrix $\mathbf{P} \propto \partial {H_{4 \times 4}}/\partial \mathbf{k}$ is block-diagonal and will not couple the up and down branches of the edge states belonging to the spin-up and spin-down families; therefore, the transition between the up and down branches is forbidden because of the conservation of the angular moment. So the experimentally measurable conductance $G(E_{F})$ is simply proportional to the total transmission probability $T_{total}\equiv T_{up}+T_{down}$ and we can define the spin polarization in the $z$ direction as: ${P_z} = ({T_{up}} - {T_{down}})/({T_{up}} + {T_{down}})$. We consider a $W=300$ nm wide Hall bar under a single magnetic barrier modulation. The parameters for the HgTe/CdTe quantum well are $A=364.5$ meV$\cdot$nm, $B=-686$ meV$\cdot$nm$^{2}$, $C=0$, $D=-512$ meV$\cdot$nm$^{2}$, and $M=-10$ meV.\cite{Bernevig} In this case, the energy spectrum of the lead has a finite-size\cite{Zhang2011} mini-gap near $E_{\mathrm{gap}}=7.3$ meV.

\begin{figure}[tbhp]
\centering
\includegraphics [width=0.95\columnwidth]{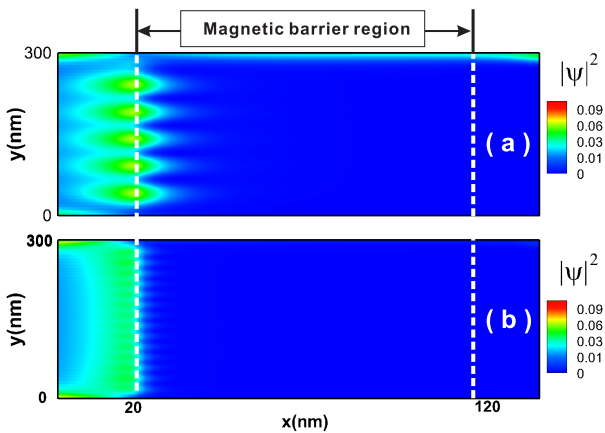}
\caption{(color online) The density distribution ${\left| {{\psi _ \uparrow }(x,y)} \right|^2}$ of the electron edge state through
the Hall bar with fixed magnetic field (a) $B=1T$ and (b) $B=2T$, respectively. The length of magnetic barrier
and the Fermi energy are fixed at $L=100$nm and $E_F=-5$ meV, respectively.}
\label{fig:3}
\end{figure}
To begin with, we illustrate the effect of $\mathbf{A}$ = $(0,A_y,0)$ without the Zeeman term $\mathbf{H_Z}$. We plot the edge states transmission probabilities ${T=T_{up}=T_{down}}$ as a function of the magnetic field for a delta-function-shaped magnetic barrier with fixed the Fermi energy in Fig. \ref{fig:2}(a). The magnetic field $B>0$ [$B<0$] corresponds to the delta-function-shaped magnetic barrier with the vector potential is depicted in Fig. \ref{fig:model}(b) [Fig. \ref{fig:model}(c)]. We find that the transmission probability ${T_{up}}$ is exactly the same as ${T_{down}}$ without the Zeeman term. The transmission ${T}$ is symmetric with respect to the different direction of the magnetic field with the same amplitudes, as shown in Fig. \ref{fig:2}(a). When the magnetic field amplitude ($\left| B \right|$) is low, the edge states transmission probabilities are perfect, i.e., $T=1$. The transmission shows a series of oscillation with increasing the magnetic field. The oscillation of the transmission probability arises from the quantum interference between the transmitted and reflected electrons in the edge states. When the magnetic field amplitude is further increased and exceeds a critical value, electrons would be blocked, and the edge channel would be switched off. For $E_F=-5$ meV, the transmission will decay slowly which results from the strong Lorentz force
with increasing magnetic field and electrons will be totally reflected back to the left side if the Lorentz force is strong enough; however, for $E_F=10$ meV, the transmitted electrons are in the evanescent modes which means the transmission will decay exponentially along the propagating direction, that's because the Fermi energy is tuned into the energy gap by increasing the magnetic field (as investigated in Ref. 25). So there are two different physical mechanisms to switch off the system.

Fig. \ref{fig:2}(b) shows the edge sates transmission probabilities as a function of the Fermi energy $E_F$ with fixed the magnetic field amplitude $B$ and the length of the magnetic barrier $L$. The sudden block of the edge state transmission around $E_F$= 7.3 meV originates from the finite-size gap: at $E_F$ =7.3 meV [shaded area in Fig. \ref{fig:2}(b), marked by $E_{gap}$]. The transmission shows oscillations as increasing the Fermi energy and the oscillation is sensitive to magnetic field amplitude. When the magnetic field is low, the transmission probability oscillates slightly and electrons propagate through the magnetic barrier perfectly away from the finite-size gap $E_{gap}$, corresponding to a well quantized conductance $G = 2e^2/h$. As increasing the magnetic field $B$, the oscillations become sharp. It's because the interference will be enhanced due to the strong Lorentz force in high-magnetic-field mechanism. The edge states transmission will be blocked completely for a certain region of incident energy $E_F$ for the high magnetic field [see the blue line in Fig. \ref{fig:2}(b)], i.e., $10$ meV $< E_F <$ $20$ meV. In the magnetic modulation region, the energy spectrum is different from that without a magnetic field as investigated Ref. 25. The transmitted electrons are in the evanescent modes and the transmission decay exponentially along the propagating direction. Consequently, it provides us with another way to control the edge states just by adjusting the Fermi energy.

In order to show the switching behavior more clearly, we plot the density distribution ${\left| {{\psi _ \uparrow }(x,y)} \right|^2}$ of the spin-up edge state at the partly or totally switch-off in Fig. \ref{fig:3}.
The magnetic field affects the coupling between the edge states and the bulk states, which leads to the backscattering process.
When the magnetic field is weak, where the system is in the switch-on state  [see Fig. \ref{fig:3}(a)], electrons can partly propagate through the Hall bar. In the strong magnetic field case, where the system is in the swith-off state [see Fig. \ref{fig:3}(b)], electrons are blocked and totally reflected back to opposite edge of the Hall bar.

\begin{figure}[tbhp]
\centering
\includegraphics [width=0.95\columnwidth]{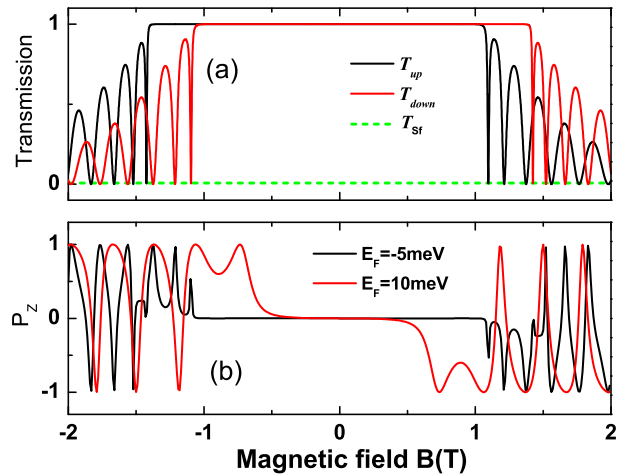}
\caption{(color online) (a) The dependence of the edge states transmission probabilities ${T_{up}}$ and ${T_{down}}$
on the magnetic field $B$ including the Zeeman term $\mathbf{H_Z}$ with $L=50$ nm, $E_F=$-5 meV.
(b) The spin polarization ${P_z}$ as a function magnetic field for different Fermi energies with $L=50$ nm.}
\label{fig:4}
\end{figure}
Next, we turn to discuss the Hall bar structure with a delta-function-shaped magnetic barrier  including the Zeeman term $\mathbf{H_Z}$. In Fig. \ref{fig:4}(a), we plot the edge state transmission probabilities ${T_{up}}$ and ${T_{down}}$ as a function of the magnetic field with the fixed Fermi energy $E_F=$ -5 meV. Notice that the Zeeman term is diagonal and wouldn't mix the up and down branches in the Hamiltonian matrix. Therefore, ${T_{\mathrm{sf}}}$ is always zero[see the green dashed lines in Fig. \ref{fig:4}(a)]. The transmission probability ${T_{up}}$ becomes different from ${T_{down}}$ when including the Zeeman term. The energy shift between ${T_{up}}$ and ${T_{down}}$ is caused by the Zeeman spin-splitting. As demonstrated in Fig. \ref{fig:4}(a), the edge state transmission ${T_{up}}$ (${T_{down}}$) is nearly perfect $T \approx 1$ while the transmission ${T_{down}}$ (${T_{up}}$) decay exponentially. It means one can make the spin-up (spin-down) edges state perfectly tunnelling through the magnetic modulation region while blocking the spin-down (spin-up) edge channel. In Fig. \ref{fig:4}(b), the spin polarization is plotted against the magnetic field for $E_F=$ -5 meV and $E_F=$ 10 meV. One can see that the spin polarization transport in the edge states can be realized by tuning the magnetic field amplitude or simply reversing the magnetization directions of the ferromagnetic strips.
The spin polarization behaviors would be distinct for different Fermi energy, which is clearly reflected in Fig. \ref{fig:4}(b).
This allows the realization of a spin-filter device using the TESs in a realistic experimental setup.

In summary, we investigated theoretically the electron tunnelling through a HgTe Hall bar with a single magnetic barrier modulation. We find (i) the edge state transmission will be suppressed by the magnetic field, (ii) the edge channel can be switched on/off by appropriately tuning the magnetic field or Fermi energy, (iii) the perpendicular magnetic field wouldn't lead to spin-flip transport, and (iv) the current can be transited from spin unpolarized to spin polarized by tuning the Fermi energy and magnetic field. These features offer us an efficient way to control the topological edge state transport, and pave a way to construct the edge state electronic device.


\end{document}